\pdfoutput=1
\documentclass[reprint, aps, prd, preprintnumbers, amsmath, amssymb, superscriptaddress, nofootinbib]{revtex4-1}
\usepackage{amssymb,amsmath,bbold,slashed}
\usepackage{graphicx,soul}
\usepackage[usenames,dvipsnames]{xcolor}
\usepackage[unicode=true,pdfusetitle,
 bookmarks=true,bookmarksnumbered=false,bookmarksopen=false,citecolor=Turquoise,
 breaklinks=false,pdfborder={0 0 1},backref=false,colorlinks=true,pdfpagemode=FullScreen]
 {hyperref}

\begin{document}

\preprint{IPMU20-0012}

\title{Upper Limit on the Proton Lifetime in Minimal Supersymetric SU(5)}

\author{Jason L. Evans}
\email[]{jlevans@sjtu.edu.cn}
\affiliation{T. D. Lee Institute and School of Physics and Astronomy, Shanghai Jiao Tong University, Shanghai 200240, China}

\author{Tsutomu T. Yanagida}
\affiliation{T. D. Lee Institute and School of Physics and Astronomy, Shanghai Jiao Tong University, Shanghai 200240, China}
\affiliation{Kavli IPMU (WPI), UTIAS, University of Tokyo, Kashiwa, Chiba 277-8583, Japan}

\begin{abstract}
In minimal supersymmetric SU(5) models, the proton can decay through dimension 5 operators. Since this decay depends directly on the supersymmetric soft masses, it will be constrained by other observables which depend on the soft masses, such as the Higgs mass and the dark matter relic density. In this work, we will examine the upper limit on the proton lifetime in minimal supersymmetric SU(5) with constrained minimal supersymmetric (CMSSM) boundary conditions set at the grand unification scale. We perform a random scan over the variables of the (CMSSM),  with of order $10^{6}$ points, and find that the proton lifetime is within reach of JUNO and Hyper-Kamiokande's experiment. 

\end{abstract}

\maketitle

\section{Introduction}
One of the compelling aspects of supersymmetry (SUSY) is that it predicts a rather precise unification of the gauge couplings. This precise unification suggests that the standard model (SM) fields should be embedded into some larger gauge group. This idea of unifying the fields and forces is further supported by the fact that when these fields are embedded into representations of SU(5), for example, we automatically get charge quantization. 

In addition to gauge coupling unification, SUSY gives a natural electroweak scale, a 125 GeV Higgs mass, and a dark matter candidate. Each of these achievements of SUSY is noteworthy in its own right. When they are combined with grand unified theories, it is hard to ignore the attraction of these models. 

The main difficulty for grand unified theories is detectability. Since the breaking scale of these theories is so high, $10^{16}$ GeV, they strongly decouple from the SM. This makes detection quite difficult. However, the requisite nature of grand unified theories, mixing of quarks and leptons, opens a window for detection. In fact, theses blurred lines between quarks and leptons leads to proton decay, the characteristic signal of all  grand unified theories. With the unified theory breaking scale so large, it might seem impossible for experiment to see such a small effect. However, because of the astounding accuracy with which experiment can measure the proton lifetime, $\sim 10^{34}$ yrs, experiments have already begun to probe the range of proton lifetimes predicted by grand unified theories. In fact, models with very strict definitions of naturalness have already been ruled out\cite{mp}. With several experiments on the horizon looking for proton decay, even more of the grand unified theories parameter will be probed. 

The goal of this work is to characterize the reach of future proton decay experiments in grand unified theories. Since a completely exhaustive study of all grand unified theories is impossible here, we limit our self to the minimal supersymmetric SU(5). Furthermore, because the dominant proton decay mode for minimal supersymmetric SU(5) is via dimension-5 operators\cite{Sakai:1981pk,Weinberg:1981wj}, the lifetime also depends on the soft SUSY breaking spectrum. The important soft parameters for proton decay are the Wino, slepton, and squark masses. Since the proton lifetime depends most strongly on the over scale of these soft masses, the details of the mass spectrum are not so important. This means that a study of the constrained minimal supersymmetric standard model (CMSSM) should be sufficient for understanding the experimental prospects of detecting proton decay more generally.  

In this work, we examine proton decay in the minimal supersymmetric SU(5) with a CMSSM SUSY breaking pattern\footnote{For studies on proton decay in other types of models see \cite{Evans:2015bxa,Ellis:2017djk,Evans:2019oyw,Evans:2020fmh,Ellis:2020mno,Ellis:2021fhb}.}. We will apply constraints coming from the Higgs mass measurement and the thermal relic dark matter density. Both of these will put an upper limit on the soft masses which will in turn put an upper limit on the proton lifetime. As we will see, this will place nearly all of the acceptable CMSSM parameter space within reach of JUNO \cite{JUNO:2015zny} and Hyper-Kamiokande \cite{Hyper-Kamiokande:2018ofw}.


\section{The Model}
Before getting to the main event, we give some background on the model we consider.  We take minimal supersymmetric SU(5), with the superpotential 
\begin{align}
 W_5 &=  \mu_\Sigma {\rm Tr}\Sigma^2 + \frac{1}{6} \lambda^\prime {\rm
 Tr} \Sigma^3 + \mu_H \overline{H} H + \lambda \overline{H} \Sigma H 
\nonumber \\
&+ \left(h_{\bf 10}\right)_{ij} 
 {\bf 10}_i {\bf 10}_j H +
 \left(h_{\bf 5}\right)_{ij} {\bf 10}_i \overline{\bf 5}_{j}
 \overline{H} ~,
\label{W5}
\end{align}
where we have suppressed the SU(5) indices, $H$ and $\bar H$ contain the SM Higgs bosons, $\Sigma$ is responsible for breaking SU(5)$\to$SU(3)$\times$SU(3)$\times$U(1), $10_i=\{Q_i,\bar U_i, \bar E_i\}$, and $\bar 5_i =\{\bar D_i, L_i\}$.

The normalization of the SU(5) breaking vacuum expectation value (vev) is $\langle \Sigma \rangle =V\cdot {\rm diag}(2,2,2,-3,-3)$ with $V=4\mu_\Sigma/\lambda'$. The GUT gauge bosons acquire a mass $M_X=5g_5 V$. The mass of the triplet Higgs boson is $M_{H_C}=5\lambda V$ and the color octet and weak triplet of $\Sigma$ get a mass $M_\Sigma=5\lambda' V/2$.  Since these masses break the SU(5) relations, they deflect the renormalization group running of the SM gauge couplings. Because the SM gauge couplings do not perfectly unify, some deflection is needed to get unification of the gauge couplings. The deflection generated by $M_{H_C}$, $M_\Sigma$, and $M_X$ masses can be adjusted to give viable gauge coupling unification. However, the colored Higgs mass, $M_{H_C}$, determined by this procedure is generally too small and leads to a proton lifetime which is too short. 

The Higgs mass predicted by this procedure is only problematic if we ignore higher dimensional operators. For example, Planck suppressed operators, which should have been included in the first place, can make the proton lifetime sufficiently long. In our analysis, we include the most important Planck suppressed operator for proton decay, which is
\begin{eqnarray}
W_{\rm eff}^{\Delta g} = \frac{d}{M_P} {\rm Tr}\left[
\Sigma {\cal W} {\cal W}
\right] ~,
\label{eq:SigmaWW}
\end{eqnarray}
With this operator included, the gauge coupling matching conditions become\cite{Evans:2015bxa,Tobe:2003yj,Hisano:1993zu} 

\begin{align}
 \frac{1}{g_1^2(Q)}&=\frac{1}{g_5^2(Q)} \\
 \nonumber & +\frac{1}{8\pi^2}\biggl[
\frac{2}{5}
\ln \frac{Q}{M_{H_C}}-10\ln\frac{Q}{M_X}
\biggr]+ (-1)\frac{8dV}{M_P}
~, \label{eq:matchg1} \\
 \frac{1}{g_2^2(Q)} & =\frac{1}{g_5^2(Q)} \\ \nonumber & +\frac{1}{8\pi^2}\biggl[
2\ln \frac{Q}{M_\Sigma}-6\ln\frac{Q}{M_X}
\biggr]+(-3)\frac{8dV}{M_P} 
~, \\
 \frac{1}{g_3^2(Q)}& =\frac{1}{g_5^2(Q)} \\ \nonumber & +\frac{1}{8\pi^2}\biggl[
\ln \frac{Q}{M_{H_C}}+3\ln \frac{Q}{M_\Sigma}-4\ln\frac{Q}{M_X}
\biggr]+(2)\frac{8dV}{M_P} ~,
\end{align}
From these expressions it is clear that the operator in Eq. (\ref{eq:SigmaWW}) leads to a splitting in the gauge couplings. Furthermore, this splitting does not diminish as we evolve the couplings up to scales beyond the GUT scale.. Because of the group theoretic coefficients in the above equations, the colored and weakly interacting subsets of SU(5) will have different effective couplings of 
\begin{eqnarray}
\frac{1}{g_3^2}-\frac{1}{g_2^2}\gtrsim 5 \epsilon
\end{eqnarray}
for all scales, where 
\begin{eqnarray}
\epsilon= \frac{8dV}{M_P}~.
\end{eqnarray}
The quantity $5\epsilon$ should be small. Otherwise, the unification of the SM gauge couplings is explicitly broken for all scales and the concept of unification becomes meaningless. Below, we will find that $5\epsilon < 0.1$ is satisfied for all points ensuring that the gauge couplings unify to a relatively high level of precision.   

We now simplify the above equations so we can understand the constraints they enforce on the masses, 
\begin{align}
 \frac{3}{g_2^2(Q)}& - \frac{2}{g_3^2(Q)} -\frac{1}{g_1^2(Q)}
\\ \nonumber &=-\frac{3}{10\pi^2} \ln \left(\frac{Q}{M_{H_C}}\right)
-\frac{96d V}{M_P} \label{eq:matchMHC}
~, \\[3pt]
 \frac{5}{g_1^2(Q)}& -\frac{3}{g_2^2(Q)} -\frac{2}{g_3^2(Q)}
\\ \nonumber &= -\frac{3}{2\pi^2}\ln\left(\frac{Q^3}{M_X^2 M_\Sigma}\right) ~,\label{eq:matchMG}
\\[3pt]
 \frac{5}{g_1^2(Q)}& +\frac{3}{g_2^2(Q)} -\frac{2}{g_3^2(Q)} \\ \nonumber &= -\frac{15}{2\pi^2} \ln\left(\frac{Q}{M_X}\right) + \frac{6}{g_5^2(Q)} -\frac{144d V}{M_P} ~. \label{eq:matchg5}
\end{align}
As is clear from these expressions, if $d=0$, then the colored Higgs mass is completely determined by the gauge couplings of the SM. As we will see below, this leads to a relatively light colored Higgs mass and so a short proton lifetime. 

Before moving on to discuss proton decay, we need to discuss the Yukawa couplings of the SM, since they play a roll in proton decay. The SM Yukawa couplings, when run up to the GUT scale, do not unify as is required in minimal supersymmetric SU(5). Although problematic, this issues is alleviated by the addition of Planck suppressed operators\cite{Ellis:1979fg,Panagiotakopoulos:1984wf,Nath:1996qs,Nath:1996ft,Bajc:2002pg},
\begin{eqnarray}
\Delta W_{Yukawa}=\frac{c_{10_{ij}}}{M_P} \Sigma {\bf 10}_i{\bf 10}_j H +\frac{c_{5_{ij}}}{M_P} \Sigma \overline{\bf 5}_i{\bf 10}_j \overline{H}~.\label{eq:YukPlanck}
\end{eqnarray}
These operators allow us to justify the lack of unification in the gauge couplings below the GUT scale.  However, it introduces an ambiguity in the value of the Yukawa couplings in the grand unified theory. Since we have no information about the coefficients $c_{5,10}$, we do not know which SM Yukawa coupling matches onto the grand unified theory's Yukawa couplings. For example, if the bottom and tau Yukawa couplings are RG run from the weak scale up to the GUT scale, we find they have different values. The Planck suppressed operators in Eq. (\ref{eq:YukPlanck}) rectify this disparity. However, we do not know which Yukawa coupling should correspond to $h_{5}$. Since some of the colored Higgs Bosons' interactions are governed by $h_5$, this ambiguity shows up in the proton decay calculation. As was shown in previous work \cite{evno}, identifying $h_5$ with the down-type quark Yukawa couplings gives a much longer lifetime. Since we are seeking an upper bound on the proton lifetime, this is what we will use here.

\section{The Proton Lifetime}
In this section, we briefly review the calculation of the proton lifetime. In matching the grand unified theory onto the minimal supersymmetric standard model (MSSM), dimension-5 operators which facilitate proton decay are generated \cite{Sakai:1981pk,Weinberg:1981wj},
\begin{eqnarray}
W_{pdecay} \!\!\!\!\!\!\!\!\!\!\!\! & =  \frac{1}{2}C_{5L}^{ijkl} \epsilon_{abc} (Q^a_i\cdot Q^b_j) (Q^c_k \cdot L_{\ell})\\ & \nonumber \quad \quad \quad \quad \quad \quad \quad \quad \quad 
+C_{5R}^{ijkl}\epsilon^{abc} (\bar{u}_{ia} \bar{e}_j \bar{u}_{kb} \bar{d}_{l c}) \, ,
\end{eqnarray}
with\footnote{We have ignored the contribution to these Wilson coefficients coming from Planck Suppressed operators. However, we take the perspective that the same mechanism that solves the Yukawa couplings hierarchy also suppresses these operators \cite{Nomura:2000yk}  } 
\begin{eqnarray}
&\!\!\!\!\!\!\! C_{5L}^{ijkl}=
\frac{\sqrt{8}}{M_{H_C}}h_{10,i}e^{i\phi_i}\delta^{ij} V_{kl}^*h_{\bar 5,l}, \\ 
&C_{5R}^{ijkl} =
\frac{\sqrt{8}}{M_{H_C}}h_{10,i} V_{ij}V^*_{kl}h_{\bar 5,l}e^{-i\phi_k} \, .
\label{eq:C5L_R_matching}
\end{eqnarray}
These operators are then run down to the SUSY breaking scale where the supersymmetric particles are integrated out generating the dimension-6 operators that govern proton decay. The resultant scattering amplitude is
\begin{align}
     {\cal A}(p\to  K^+ &\bar{\nu}_i)= \\
C_{LL_i}&\left(\langle K^+\vert (us)_Ld_L\vert p\rangle
\langle K^+\vert (ud)_Ls_L\vert p\rangle\right) \nonumber 
 \\
+C_{RL_1}&\langle K^+\vert (us)_Rd_L\vert p\rangle
+ C_{RL_2}\langle K^+\vert (ud)_Rs_L\vert p\rangle  \nonumber ~.
\end{align}
where a rough approximation for the coefficients ,which exhibits their scaling with different parameters, is \cite{Ellis:2019fwf}
\begin{align}
 C_{LL_i}\simeq
\frac{2\alpha_2^2}{\sin 2\beta}&\frac{m_t m_{d_i} M_2}{ m_W^2M_{H_C}M_{\rm
 SUSY}^2} V_{ui}^*V_{td}V_{ts}e^{i\phi_3}\\
 \times \nonumber &\left(1 +
 e^{i(\phi_2-\phi_3)}\frac{m_c V_{cd}V_{cs}}{m_tV_{td}V_{ts}}\right) ~,
\label{eq:cllaprox}
\end{align}
and
\begin{align}
 C_{RL_1} &\simeq -
 \frac{\alpha_2^2}{\sin^2 2\beta}\frac{m_t^2 m_s m_\tau \mu}{ m_W^4 M_{H_C}M_{\rm SUSY}^2}V_{tb}^*
 V_{us} V_{td}e^{i\phi_1} ~, \\
 C_{RL_2} &\simeq -
 \frac{\alpha_2^2}{\sin^2 2\beta}\frac{m_t^2 m_d m_\tau \mu}{m_W^4 M_{H_C} M_{\rm SUSY}^2}V_{tb}^*
 V_{ud} V_{ts}e^{i\phi_1} ~,
\label{eq:crlaprox}
\end{align}
where $\alpha_2=g_2^2/4/\pi$, $m_W$ is the W boson mass, $V_{ij}$ are the CKM matrix elements, $\beta$ is the measure of the Higgs mass ratio, $M_{H_C}$ is the colored Higgs mass, $\phi_i$ are GUT scale phases from the Yukawa couplings \cite{Ellis:1979hy}, $M_2$ is the Wino mass, $M_{SUSY}$ is the average mass scale of the soft mass, $\mu$ is the Higgs bilinear mass, $(m_t,m_{d_i},m_s)$ are the quark masses, and $m_\tau$ is the tau mass.

The proton decay width is then found to be 
\begin{equation}
 \Gamma(p\to K^+\bar{\nu}_i)
=\frac{m_p}{32\pi}\biggl(1-\frac{m_K^2}{m_p^2}\biggr)^2
\vert {\cal A}(p\to K^+\bar{\nu}_i)\vert^2~,
\end{equation}
where $m_p$ is the proton mass and $m_K$ is the Kaon mass\footnote{ For more details on the calculation of the proton lifetime see \cite{eelnos}}.


\section{Results}
In this section, we present the results of our scan over of order $ 10^{6}$ points with CMSSM like boundary masses in the ranges
\begin{eqnarray}
& m_{1/2} \; \in [2000, 20000]~{\rm GeV}\, , \\  & m_0 \; \in [2000, 20000]~{\rm GeV}\, , \\
&A_0/m_0 \; \in [-3.5, 3.5],\quad  \lambda \; \in [0, 1] \, , \\ 
&\lambda' \; \in [0, 10^{-3}], \quad  \tan \beta \; \in [2, 20]\, , &
\end{eqnarray}
with a linear measure. As is seen above, we do not scan over the phase of the Yukawa couplings, $h_5,h_{10}$. Instead, we search for the phase with the maximum lifetime and use that phase in our lifetime calculation. Since the phases only effect is on the proton lifetime, we think this is justified in a search for the maximum lifetime. Another important point of the above equations is we do not scan over $d$ and do scan over $\lambda$ and $\lambda'$. This is because the matching conditions in Eq. (\ref{eq:matchMHC})-(\ref{eq:matchg5})) give us three conditions on the five parameters $(V,\lambda,\lambda',g_5,d)$. We can choose any two as independent parameters. We have chosen $\lambda,\lambda'$ as independent and $d$ as determined. We make this choice since it leads to a more straight forward relationship between $\lambda,\lambda'$ and  $M_{H_C}$. With this choice, the colored Higgs mass scales roughly as $\lambda/\lambda'^{1/3}$ \footnote{This can be determined using Eq. (\ref{eq:matchMHC})-(\ref{eq:matchg5})) and the relations between the masses and the vev.}. This means smaller values of $\lambda'$ will give larger values of $M_{H_C}$ and so a longer proton lifetime. Since it is unlikely $M_{H_C}$ is large enough to give an acceptable proton lifetime, unless $\lambda'< 10^{-3}$, we restrict our scan to this range. Below, we will discuss some of the important factors for getting a sufficiently long proton lifetime.

\begin{figure}[t!]
\includegraphics[width=.55\textwidth]{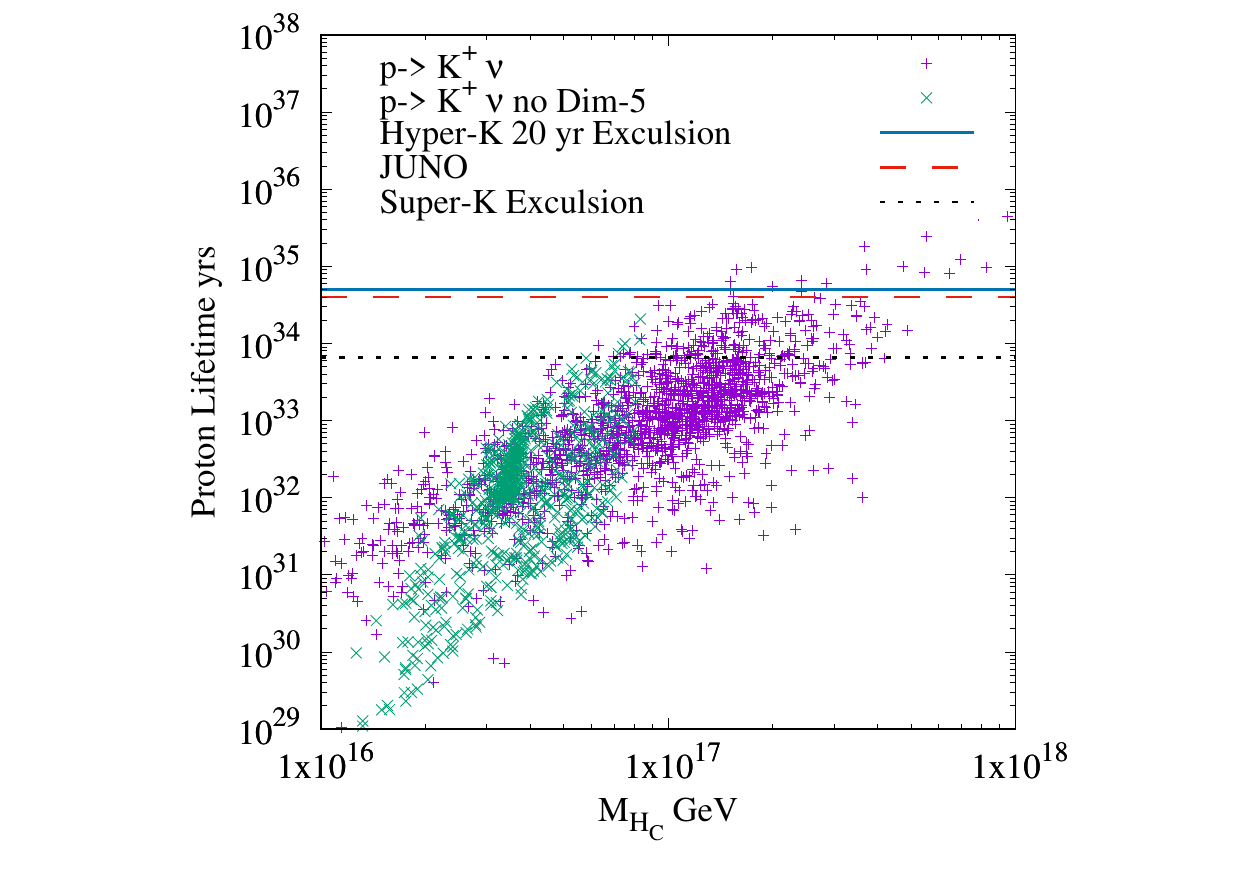}
\caption{Here we plot the proton lifetime as a function of $M_{H_C}$ for points which have a Higgs mass in the range $m_h=122-128$ GeV, a relic density $\Omega_\chi h^2 < 0.12$, and have a viable direct detection cross section. The green points are for $d=0$ and the purple points are for $d\ne 0$. The black dotted line is the current limit from Super-Kamiokande, the red dashed line is JUNO's 20 year reach, and the blue solid line is the 20 year reach of Hyper-Kamiokande. \label{fig:c5vsnoc5} }
\end{figure}

\begin{figure}[t!]
\includegraphics[width=.55\textwidth]{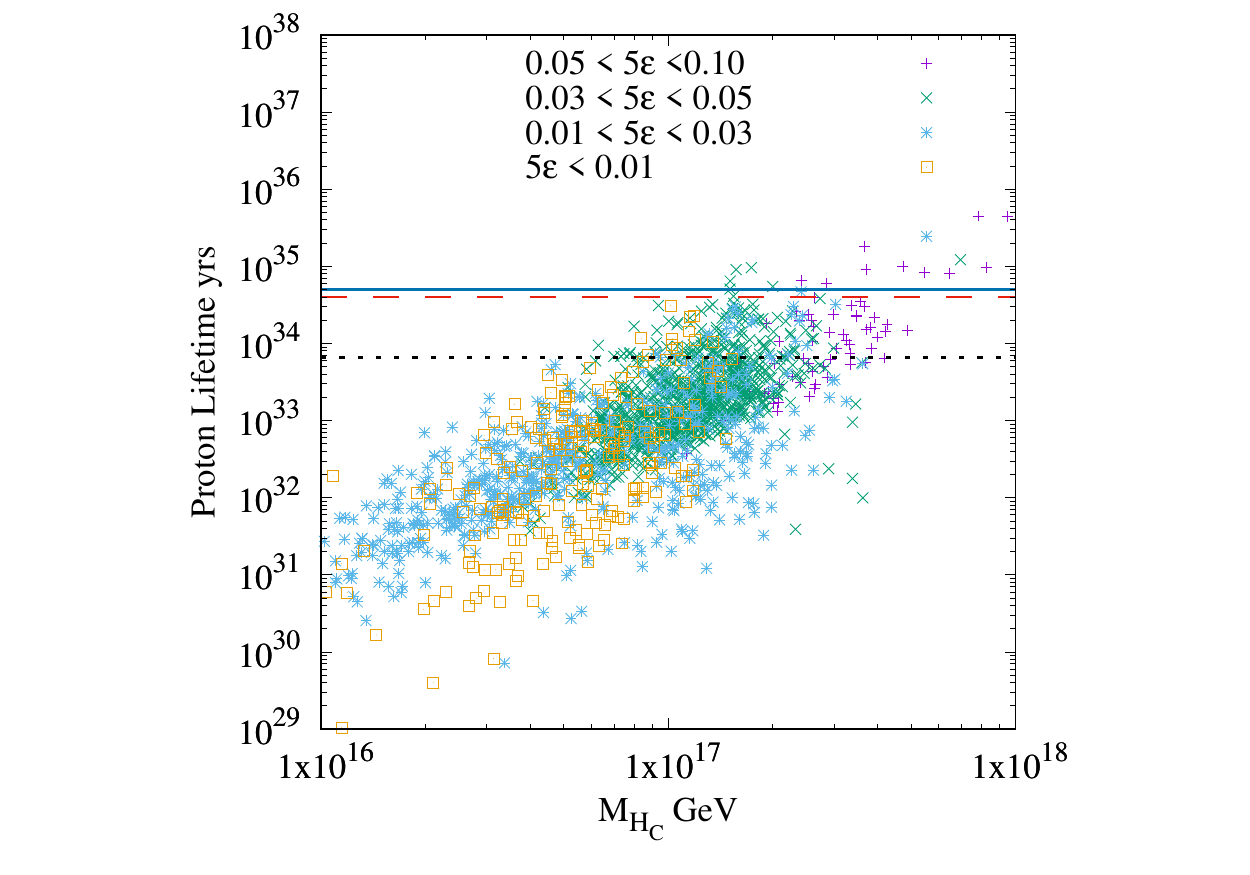}
\caption{Here we plot the proton lifetime as a function of $M_{H_C}$ for points which have $d\ne=0$, a Higgs mass in the range $m_h=122-128$ GeV, a relic density $\Omega_\chi h^2 < 0.12$, and have a viable direct detection cross section. The points are sorted according to the size of $\epsilon$. See text for discussion of $\epsilon=8dV/M_P$. The three horizontal lines are the same as in Fig. (\ref{fig:c5vsnoc5}). \label{fig:eps}  }
\end{figure}

Here we discuss the important phenomenological constraints we apply to our scans. We require each point to have a Higgs mass in the range $122-128$ GeV, which is calculated using FeynHiggs 2.18.0 \cite{FH}. We also restrict the relic density to the range $\Omega_\chi h^2< 0.12$. Furthermore,  we apply the dark matter direct detection limits using MicroOmegas \cite{Belanger:2001fz,Belanger:2004yn}.

We now report our results. In Fig. (\ref{fig:c5vsnoc5}), we plot the proton lifetime versus $M_{H_C}$. For each point, we calculate the proton lifetime for $d\ne 0$ and $d=0$. For $d=0$, the colored Higgs mass is constrained to be less than about $10^{17}$ GeV. This leads to an upper limit on the proton lifetime just beyond the Super-Kamiokande limit. This means that proton decay constraints nearly rule out minimal SU(5) with CMSSM like boundary conditions and $d=0$. As can be seen in the figure, for the points with $d\ne0$, the colored Higgs mass can be much larger and the number of points beyond the Super-Kamiokande limit is significantly increased. Interestingly, an overwhelming number of those points are within reach of JUNO and Hyper-Kamiokande. As is clear from this figure, the higher dimensional operator in Eq. (\ref{eq:SigmaWW}) has a rather important effect on the lifetime and consequence for proton decay searches. Clearly, there are great prospects for proton decay at JUNO and Hyper-Kamiokande.

From Fig. (\ref{fig:c5vsnoc5}) it is also clear that the points beyond the reach of JUNO and Hyper-Kamiokande are less good from a theoretical stand point. These points have a colored Higgs mass quite close to the Plancks scale. Although this does not rule these points out it does make them a bit suspect. Points with $M_{H_C}\gtrsim M_P$ are indeed ruled out since they would form black holes. 

\begin{figure}[t!]
\includegraphics[width=.51\textwidth]{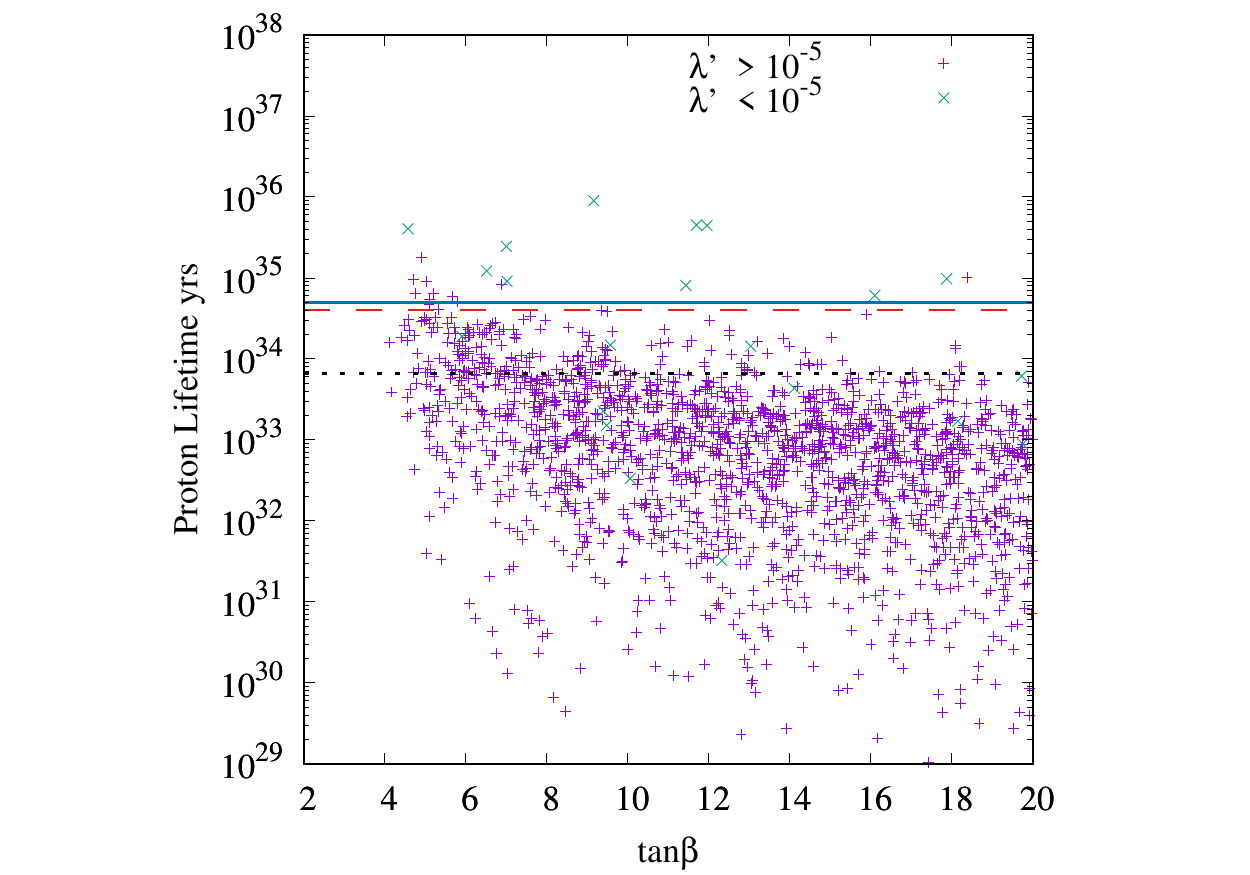}
\caption{Here we plot the proton lifetime as a function of $tan\beta$ for points which have $d\ne0$, a Higgs mass in the range $m_h=122-128$ GeV, a relic density $\Omega_\chi h^2 < 0.12$, and have a viable direct detection cross section. The three horizontal lines are the same as in Fig. (\ref{fig:c5vsnoc5}). \label{fig:tanb}  }
\end{figure}

In Fig. (\ref{fig:eps}), we again plot the lifetime against $M_{H_C}$. However, we now sort the points by the value of $\epsilon$. This gives us a rough relationship between the gauge couplings splitting and the proton lifetime. Examining Eq. (\ref{eq:matchMHC}), it is clear that generally if $\epsilon$ is larger $M_{H_C}$ is larger and, thus, so is the proton lifetime. Although we do not exclude any points based on the value of $\epsilon$ used, we do believe that points with larger values of $\epsilon$ are inferior since they are a source of the incomplete gauge coupling unification at all scale. Since the points with the largest values of $\epsilon$ tend to arise for $M_{H_C}$ close to the Planck scale, these points are problematic on multiple levels. 

In the final figure, Fig. (\ref{fig:tanb}), we present the proton lifetime versus $\tan\beta$. This figure clearly shows that the proton lifetime constraints prefer smaller values of $\tan\beta$. This preferences is due to a suppression of the Yukawa coupling relevant for proton decay when $\tan \beta$ is small.  In fact, super-Kamiokande constraints have already forced $\tan\beta\lesssim 15$. As the coming experiment accrue more data, this bound will become much more severe if no proton decay is detected. Another interesting feature of this figure is the $\lambda'$ dependence of proton decay. In this figure, we have sorted the points by whether $\lambda'$ is larger or smaller than  $10^{-5}$. As can be seen, the points with their lifetime beyond the reach of JUNO and Hyper-Kamiokande tend to have rather small $\lambda'$. These are the same points which have $M_{H_C}\sim M_P$ and larger values of $\epsilon$. All things considered, the points beyond the reach of JUNO and Hyper-Kamiokande come from points with an inferior set of parameters. 

The rough upper bound on the proton lifetime in this study can be understood by examining Eq. (\ref{eq:crlaprox}) and (\ref{eq:cllaprox}). The largest proton lifetimes come from points with large $M_{H_C}$ and $M_{SUSY}$, and  small $M_2$ and $\mu$. The need for large $M_{H_C}$ was discussed above. We now discuss the effects of the low-scale parameters.

Because of unification of the gauginos, the lightest gaugino is Bino like. This leaves us with two possible dark matter candidates, a Bino-like or Higgsino-like candidate. Bino-like dark matter is ruled out by direct detection unless we are in a regime where coannihilation occurs. This corresponds to a Bino mass larger than a TeV.  For a dominantly Higgsinos dark matter candidate, we need a TeV Higgsinos and everything else larger.  This puts constraints on how small $\mu$ and $M_2$ can be. This prevents a super long proton lifetimes. The Higgs mass measurement forces the soft masses to be not too large, again preventing a excessively long proton lifetime. Therefore, it is reasonable to conclude that all models with universal gaugino masses at the GUT scale will give similar upper bounds on the proton lifetime \footnote{One caveat is models with anomaly mediated gaugino masses \cite{Randall:1998uk,Giudice:1998xp}.} .

\section{Conclusion}
In this paper, we have addressed the detectability of minimal supersymmetric SU(5) with CMSSM boundary conditions. We scanned over grand unified theories with CMSSM like boundary masses and searched for the maximum proton lifetime among the models with other good phenomenological features, i.e. a Higgs mass consistent with 125 GeV and a relic density $\Omega_\chi h^2<0.12$. To ensure our predicted proton lifetime was indeed maximal, we included effects that come from matching the Yukawa couplings, scanning over the phase of Yukawa couplings, and higher dimensional operators.

Our above results convincingly shows that the majority of points, which have a viable Higgs mass and dark matter candidate, are within reach of JUNO and Hyper-Kamiokande. Furthermore, the points which are beyond the reach of JUNO and Hyper-Kamiokande tend to have some theoretically concerning features, e.g. $M_{H_C}\sim M_P$, large $\epsilon$, or $\lambda'\lesssim 10^{-5}$. Thus, this work has made a rather strong case for minimal supersymmetric SU(5) being within reach of the coming set of experiments if a CMSSM like spectrum is assumed\footnote{Analysis of a more generic spectrum will be give in future work \cite{ES}}.  

This result means that the coming experiments can make meaningful statements about grand unified theories whether they see proton decay or not. Furthermore, we have shown that well motivated models like the minimal supersymmetric SU(5) can still be tested strongly supporting the need for their further study.

\end{document}